\documentclass[%
 reprint,
 amsmath,amssymb,
 aps,
pra,
]{revtex4-1}
\usepackage{graphicx}
\usepackage{dcolumn}
\usepackage{bm}
\usepackage{epstopdf}
\usepackage{verbatim}
\usepackage{makecell}
\usepackage{multirow}
\usepackage{amsmath,amssymb,amsfonts}
\usepackage{xcolor}

\usepackage[dvipdfm,colorlinks,linkcolor=blue, urlcolor=blue, anchorcolor=blue,citecolor=blue]{hyperref}
\begin{document}
\maxdeadcycles=1000
\preprint{APS/123-QED}

\title{Adiabatic light propagation in nonlinear waveguide couplers with longitudinally varying detunings via resonance-locked inverse engineering}

\author{Fu-Quan Dou}
\email{doufq@nwnu.edu.cn}
\affiliation{College of Physics and Electronic Engineering, Northwest Normal University, Lanzhou, 730070, China}
\author{Ya-Ting Wei}
\affiliation{College of Physics and Electronic Engineering, Northwest Normal University, Lanzhou, 730070, China}
\author{Zhi-Ming Yan}
\affiliation{College of Physics and Electronic Engineering, Northwest Normal University, Lanzhou, 730070, China}


\begin{abstract}
We investigate the adiabatic evolution of light in nonlinear waveguide couplers via resonance-locked inverse engineering based on stimulated Raman adiabatic passage (STIRAP). The longitudinal varying detunings of the propagation coefficients are designed to eliminate dynamically the nonlinear effect, which induce the non-adiabatic oscillations. We show that different light evolutions such as complete light transfer, light split and light return can be realized adiabatically with appropriate choices of the detunings even in the nonlinear regime. The features open new opportunities for the realization of all-optical nonlinear devices with high fidelity in integrated optics.
\end{abstract}
\maketitle
\section{Introduction}\label{intro}
The waveguide (WG) couplers with tolerance to variations in fabrication and input wavelength are highly desirable with many potential and significant applications in area of optical circuits
and communications \cite{Ramadan:98,Paloczi2004,Sun:09,Chen:16}. Such devices have been widely used as switches \cite{Sekiguchi:12}, beam splitters \cite{Chenxi2018}, polarization rotators \cite{Chen:14} and other building blocks of optical waveguides \cite{PhysRevLett.101.200502,Yu2008}. They can be fabricated in the silicon-on-insulator (SOI) platform due to their simple configurations and the ease \cite{Lu:15}. To our knowledge, the most commonly used is adiabatic couplers because they do not require strict fabrication
control and are broadband \cite{Ho:15,PhysRevA.93.033802,PhysRevA.87.013806,PhysRevE.73.026607} in comparison to the conventional couplers which are known to be highly sensitive to structural parameters.

Building on the strong similarity between quantum mechanics and waveguide optics, coupled optical waveguides have proven to provide a very rich laboratory tool to investigate with optical waves the classical analogues of a wide variety of coherent quantum effects \cite{PhysRevA.71.065801} encountered in atomic, molecular or condensed-matter physics \cite{Longhi2009}. On the other hand, many coherent protocols based on quantum techniques have been exploited to manipulate light propagation in optical waveguide couplers, such as stimulated Raman adiabatic passage (STIRAP) \cite{Longhi2009,PhysRevA.70.063409,RevModPhys.89.015006}, shortcut to adiabaticity (STA) \cite{Ho:15,Tseng:14,Dou2022,Zhang_2021}, for instance, counter-diabatic driving (or quantum transitionless driving) \cite{Chenxi2018,Demirplak2003,Demirplak2008,PhysRevA.90.060301}, dressed-state technique \cite{PhysRevLett.116.230503,Dou2020}, adiabatic elimination \cite{PhysRevA.97.023811,PhysRevA.29.1438}, etc. These methods are committed to the exploitment of devices with high coupling or conversion efficiency at specific device lengths which feature good robustness against wavelength and fabrication variations \cite{Ho:15}. However, realistic propagation of light in the waveguide which is associated with nonlinear optical process \cite{PhysRevLett.101.193901}. When the nonlinearity come into play in waveguides, using these tools is a challenging question. At the same time, the general instability of the dynamics prevents the direct application of simple strategies \cite{PhysRevA.89.012123,PhysRevA.98.022102,PhysRevA.93.043419,PhysRevA.78.053410,PhysRevA.81.052112}.

The study of waveguide couplers that contain nonlinear materials have triggered great interest in the integrated optics \cite{Dang2017,Julius2022,Hanapi2021,Deka2018,Jia:10,Schiek2012,PhysRevB.88.045443,Goto2016} and are critically essential for the implementation in future quantum computers \cite{Goto2016} due to their novel all-optical switching applications \cite{Schiek2012,PhysRevB.88.045443} since being introduced by Jensen in 1982 \cite{Jensen1982}. It has been demonstrated that light propagation (light transfer, light split, light return, etc.) in waveguide systems exhibits intriguing features in the presence of optical nonlinearity \cite{Jia:10,Diebel2014,Friberg1987,deLima2018,Venugopal2012,Petr2016}. The nonlinear couplers can exhibit self-trapping, self-modulation and self-switching of the energy of the coupled modes \cite{Leoski2004}. For the analysis of problems of light pulse propagation in nonlinear optical couplers, the coupled-mode theory (CMT) \cite{Huang1994,Yuan1994} is also adopted which provides an easy and intuitive way for us to obtain physical insight. The important and natural phenomenon of optical fibers is the Kerr (third-order) nonlinearities due to the change of the refractive index of partial coupler structure as a result of high propagating power \cite{Zhang2021,PhysRevA.91.013840,Dacles-Mariani2007, Qi2014,Wang2004,Fleischer:05,PhysRevA.93.023848,PhysRevE.82.056605}. Of course, the question naturally arises on how nonlinear effects may influence such adiabatic transfer processes \cite{PhysRevLett.101.193901,PhysRevA.73.013617}. They are known to induce dynamically unstable regions in the parameter spaces \cite{PhysRevLett.93.250403,PhysRevA.73.013617,PhysRevLett.119.243902,PhysRevA.72.013608,PhysRevA.93.023848,PhysRevE.82.056605} and impair the efficiency of STIRAP drill on theory and experiment \cite{PhysRevLett.101.193901}. This failure was explained by the destruction of the dark state formed in the STIRAP configuration \cite{PhysRevLett.93.250403,PhysRevA.72.013608,PhysRevLett.101.193901,PhysRevLett.119.243902,PhysRevA.73.013617}. To overcome this detriment of  nonlinearity, recent work established an exact resonance-locked inverse engineering technique allows one to induce a controlled population inversion which surpasses the usual nonlinear STIRAP efficiency in the case of quantum $\Lambda$ systems featuring second- and third-order nonlinearities \cite{PhysRevLett.119.243902}.

In this paper, we theoretically propose a design of the nonlinear waveguide couplers by the resonance-locked inverse engineering based on the STIRAP. The resonance-locked inverse protocol shows how the efficiency of the nonlinear STIRAP can be improved by the longitudinal varying detunings of the propagation constants depending on the desired evolution of the light intensity, which aims to nullify the nonlinear effect. We also demonstrate that our devices could achieve complete energy transfer, beam splitting and light return with robustness against coupling parameters fluctuations.

This paper is organized as follows. In Sec. \ref{sec:2}, we present the Hamiltonian of the nonlinear waveguide optical model and briefly explain the basic idea of the resonance-locked inverse engineering. The numerical results for the spatial evolution of the three light distributions in the single-mode waveguides for different coupling constants under study are reported in Sec. \ref{sec:3}. Finally, the conclusions are given in Sec. \ref{sec:4}.
\section{Model and approach}\label{sec:2}
For an array of $N+1$ optical waveguides with Kerr nonlinearity, in the framework of the CMT, the evolution of the wave amplitude with nearest-neighbor evanescent coupling is accurately described by a set of $N+1$ coupled differential equations \cite{Longhi2009}
\begin{equation}
\label{eq1}
i\frac{d}{dz}\mathbf{a}(z)=H(z)\mathbf{a}(z),
\end{equation}
where $\mathbf{a}(z)=[a_{1}(z),a_{2}(z),\ldots, a_{N+1}(z)]^T$ is the electric field amplitude of the individual waveguides. Replacing the spatial variation $z$ with the temporal variation $t$, Eq. (\ref{eq1}) is equivalent to the time-dependent Schr\"{o}dinger equation ($\hbar=1$). The $H(z)$ describes the interaction between the waveguide modes could be
explicitly read as (in matrix form) \cite{Christodoulides1988,Sukhorukov2003}
\begin{widetext}
\begin{equation}
\setlength{\arraycolsep}{2pt}
H(z)=\left[\begin{array}{ccccccc}
     \beta_{1}+\Gamma|a_{1}(z)|^2 & C_{1} &0 &\cdots&0\\
     C_{1}&\beta_{2}+\Gamma|a_{2}(z)|^2&C_{2}&\cdots&0\\
     0 & C_{2} & \ddots &\ddots&\vdots\\
     0&0 & \ddots & \ddots & C_{N}\\
     0&0 & \cdots & C_{N} & \beta_{N+1}+\Gamma|a_{N+1}(z)|^2
\end{array}\right],\label{eq2}
\end{equation}
\end{widetext}
where the $C_{k}$ $(k=1,2,3,\cdots,N)$ is the $z$-dependent coupling coefficients which can be achieved with changing the distance between the waveguides. The dynamics preserves the normalization, which is fixed as $\sum_{i=1}^{N+1}|a_{i}(z)|^2=1$. $\beta_{i}$ denotes the linear propagation constant. It is assumed that the refractive index $n$ of the material is intensity-dependent, i.e., $n = n_{0} + n_{2}|E|^2$ \cite{PhysRevA.91.013840,Dacles-Mariani2007}. Such a dependence of $n$ results in linear term $\beta_{i}$ and nonlinear term $\Gamma|a_{i}(z)|^2$ in Eq. (\ref{eq1}). $\Gamma$ characterizes the Kerr nonlinearity which may lead to dynamical instability \cite{PhysRevLett.101.193901,PhysRevA.73.013617,PhysRevLett.93.250403,PhysRevLett.119.243902,PhysRevA.72.013608,PhysRevA.102.052203,PhysRevLett.98.050406,PhysRevLett.121.250405,Tsoy2018}. In the waveguide system, nonlinearity is introduced by increasing the power of the input beam \cite{PhysRevLett.81.3383,PhysRevLett.95.073902,PhysRevLett.74.2941}. At low light power levels, $\Gamma$ can be neglected ($\Gamma=0$) and the corresponding system is perfectly analogous to the linear case. With the emergence of the nonlinearity, the differential equation Eq. (\ref{eq1}) is no
longer analytically solvable, we therefore exploit a fourth-fifith order Runge-Kutta algorithm to trace the light evolution numerically. Without loss of generality, we take the coupling and nonlinear coefficients to be real and positive and the variation of the coupling coefficients due to nonlinearity at high input intensity are neglected in our works. Moreover, the study of beam dynamics in nonlinear waveguide couplers \cite{PhysRevA.93.023848,Wang2004,Fleischer:05} is of great significance for light routing in photonic circuits, for beam steering and switching \cite{Tsoy2018,RevModPhys.83.247}.

When the intensity of the optical pulses crosses a certain threshold value then the optical fiber behaves nonlinearly \cite{PhysRevLett.74.2941}. The existence of the nonlinearities make it not justified to apply the adiabatic condition of linear STIRAP to nonlinear case \cite{PhysRevLett.90.170404,PhysRevA.73.013617,MENG2009,PhysRevLett.99.223903} because of the absence of the superposition principle. More interestingly, the adiabatic
geometric phase or Berry phase is found also to be modified by the nonlinearity \cite{PhysRevA.81.052112}. There may exist more eigenstates than the dimension of the Hilbert space \cite{PhysRevLett.90.170404,MENG2009} and the non-adiabatic coupling between these eigenstates will drastically decrease the transfer efficiency \cite{PhysRevLett.99.223903,PhysRevA.73.013617}. Therefore, it is very difficult to analyze the adiabaticity of the nonlinear systems and follow their adiabatic conditions \cite{PhysRevLett.98.050406,PhysRevA.78.053410}. The key here for the success of nonlinear STIRAP is how to avoid these unstable regimes when designing the route of adiabatic passage.

We wish to nullify the nonlinear effect of nonlinear coupled-waveguide devices using resonance-locked engineering scheme \cite{PhysRevLett.119.243902} based on the STIRAP. This method is performed by introducing a general parametrization of the state vector and further derive the necessary conditions to obtain an efficient transfer. Now we employ resonance-locked inverse engineering technique to the system. We have to consider the phase $\gamma$ on parametrization
\begin{eqnarray}
a_{n}\mapsto a_{n}e^{-i\gamma},
\end{eqnarray}
this phase $\gamma$ is $z$-dependent and will be used to impose the resonance locking [see Eq. (\ref{eq6})]. Then the Eq. (\ref{eq1}) can be written as
\begin{equation}
i\frac{d}{dz}\mathbf{a}(z)=(H(z)-\dot{\gamma})\mathbf{a}(z)=H'(z)\mathbf{a}(z),
\end{equation}
and the resulting Hamiltonian $H'(z)$ becomes
\begin{widetext}
\begin{equation}
\setlength{\arraycolsep}{0.8pt}
H'(z)=\left[\begin{array}{ccccccc}
     \beta_{1}+\Gamma|a_{1}(z)|^2-\dot{\gamma} & C_{1} &0 &\cdots&0\\
     C_{1}&\beta_{2}+\Gamma|a_{2}(z)|^2-\dot{\gamma}&C_{2}&\cdots&0\\
     0 & C_{2} & \ddots &\ddots&\vdots\\
     0&0 & \ddots & \ddots & C_{N}\\
     0&0 & \cdots & C_{N} & \beta_{N+1}+\Gamma|a_{N+1}(z)|^2-\dot{\gamma}
\end{array}\right].
\end{equation}
\end{widetext}
In order to completely eliminate the nonlinear effects dynamically, where the dynamics is exactly the same as the linear case, we shilling one of the elements on the main diagonal equal to zero, i.e., $\beta_{i}+\Gamma|a_{i}(z)|^2-\dot{\gamma}=0$, and then substitute it into the other remaining main diagonal elements, which leads to the resonance-locked conditions \cite{PhysRevLett.119.243902} as follows
\begin{equation}
\label{eq6}
\beta_{j(j\neq i)}=\beta_{i}+\Gamma|a_{i}(z)|^2-\Gamma|a_{j}(z)|^2,
\end{equation}
these propagation constants $\beta_{j}$ $(j=1,2,3\cdots,N+1)$, fully compensate the Kerr terms which based on population dynamics, thus we can just recover the preceding equations without Kerr terms.
\section{Adiabatic light propagation in nonlinear waveguide couplers}\label{sec:3}
For the sake of simplicity, we consider a waveguide structure with three waveguides in a nonlinear coupling configuration. In the framework of CMT, the evolution of the light amplitudes $a_{1}(z)$, $a_{2}(z)$ and $a_{3}(z)$, in the presence of the optical Kerr effect, can be described by the set of coupled discrete nonlinear Schr\"{o}dinger equation (DNLSE) with the Hamiltonian given by,
\begin{equation}
\label{eq3}
\setlength{\arraycolsep}{1pt}
H_{0}(z)=\left[\begin{array}{ccc}
     \beta_{1}+\Gamma|a_{1}(z)|^2 & C_{1} &0\\
     C_{1}&\beta_{2}+\Gamma|a_{2}(z)|^2&C_{2}\\
     0&C_{2}& \beta_{3}+\Gamma|a_{3}(z)|^2
\end{array}\right].
\end{equation}
It is easy to show, as in the linear counterpart, the nonlinear systems also support the existence of the coherent population trapping (CPT) state or dark state with zero eigenvalue, with the corresponding state vector given by $|\Psi_{0}\rangle=(a_{1}^0,a_{2}^0,a_{3}^0)$, where $a_{1}^0=\frac{C_{2}}{\sqrt{C_{1}^2+C_{2}^2}}$, $a_{2}^0=0$ and $a_{3}^0=\frac{C_{1}}{\sqrt{C_{1}^2+C_{2}^2}}$. The existence of the CPT state cannot guarantee that it can always be followed adiabatically. The nonorthogonality between the CPT state and any of the other eigenstates is quite obvious. In order to keep the mathematics as simple as possible without sacrificing the central physics, we have set here $\dot{\gamma}=\beta_{2}+\Gamma|a_{2}|^{2}$ by the same procedures above, and the resonance-locked conditions can be obtained as
\begin{eqnarray}
\label{eq4}
\begin{aligned}
\beta_{1}=&\Gamma(|a_{2}|^{2}-|a_{1}|^{2})+\beta_{2},\\
\beta_{3}=&\Gamma(|a_{2}|^{2}-|a_{3}|^{2})+\beta_{2},
\end{aligned}
\end{eqnarray}
then the nonlinear Hamiltonian Eq. (\ref{eq3}) can be rewritten as
\begin{equation}
\setlength{\arraycolsep}{2.5pt}
H_{0}'(z)=\left[\begin{array}{ccc}
     0 & C_{1} &0\\
     C_{1}&0&C_{2}\\
     0&C_{2}& 0
\end{array}\right].
\end{equation}
Subsequently, we can calculate the the propagation matrix $H_{0}'(z)$ corresponding eigenvalues $E_{0}=0$, $E_{\pm}=\pm\sqrt{C_{1}^2+C_{2}^2}$ and the eigenstates given by
\begin{eqnarray}
\begin{aligned}
|\Phi_{0}(z)\rangle=&\cos\theta|1\rangle-\sin\theta|3\rangle,\\
|\Phi_{\pm}(z)\rangle=&\frac{1}{\sqrt{2}}\sin\theta|1\rangle \pm\frac{1}{\sqrt{2}}|2\rangle \frac{1}{\sqrt{2}}\cos\theta|3\rangle,
\end{aligned}
\end{eqnarray}
where the mixing angle $\theta$ is defined as
\begin{equation}
\tan\theta=\frac{C_{1}}{C_{2}},
\end{equation}
and the adiabatic Hamiltonian
\begin{equation}
\setlength{\arraycolsep}{2.5pt}
H_{a}(z)=\left[\begin{array}{ccc}
     \sqrt{C_{1}^2+C_{2}^2} & \frac{1}{\sqrt{2}}i\dot{\theta} &0\\
     -\frac{1}{\sqrt{2}}i\dot{\theta}&0& -\frac{1}{\sqrt{2}}i\dot{\theta}\\
     0& \frac{1}{\sqrt{2}}i\dot{\theta}&  -\sqrt{C_{1}^2+C_{2}^2}
\end{array}\right].
\end{equation}
The off-diagonal terms of $H_{a}(z)$ are regarded as a non-adiabatic correction. When the adiabatic criterion, which can be written as $\sqrt{C_{1}^2+C_{2}^2}\gg \dot{\theta}$, is satisfied, there is no transition between the adiabatic states during the evolution. Mathematically, it means that for the entire dynamical evolution, the system remains at one of the systems eigenmodes \cite{Suchowski2014}.

When the evolution of the system is adiabatic, we can continue to predict the dynamics of light propagation in waveguides which obeys the following dark state $|\Phi_{0}(z)\rangle$, light intensity contained in WG1 and WG3 can be described as $|a_{1}|^{2}=\cos^2\theta$, $|a_{3}|^{2}=\sin^2\theta$ with $|a_{2}|^{2}=0$, then propagation constants mismatch under the resonance-locked inverse method can be simplified as
\begin{eqnarray}
\begin{aligned}
\Delta\beta_{12}=-&\Gamma\cos^{2}\theta,\\
\Delta\beta_{32}=-&\Gamma\sin^{2}\theta,
\end{aligned}
\end{eqnarray}
where $\Delta\beta_{12}=\beta_{1}- \beta_{2}$ and $\Delta\beta_{32}=\beta_{3}- \beta_{2}$. In addition, the nonlinear constant $\Gamma$ is evaluated to be $\Gamma=5m^{-1}W^{-1}$ \cite{PhysRevLett.101.193901,PhysRevLett.81.3383,PhysRevLett.95.073902} in our calculations. Based on the above, we can see the originally introduced phase $\gamma$ can be simplified to $\beta_{2}z$ which is commonly used in the waveguide systems \cite{PhysRevA.97.023811,Paspalakis2006,PhysRevE.73.026607,PhysRevA.103.053705}. The validity of the method is further substantiated by the fact that, when applied to the nonlinear system, the same result is obtained as in linear case. There is still much freedom to design these propagation constants detunings at will depending on the mixing angle $\theta$ and Kerr nonlinear constant $\Gamma$. It is noted that the corresponding CPT state via STIRAP are identical in form in both linear and nonlinear system, which are indeed governed by different adiabatic conditions \cite{PhysRevLett.98.050406}. When the above two conditions are satisfied at the same time, it can completely eliminate the nonlinear effect. Indeed, the adiabatic condition for the nonlinear system is more difficult to fulfill compared with its linear counterpart which limits the efficiency of the resonance-locked inverse method. Thus the result of the resonance-locked inverse engineering is not always strictly reachable desired final states in nonlinear system.

Since there are many possibilities of the two coupling coefficients, various light evolutions in the waveguide structure can be expected. Thus we are going to investigate three possible situations in the following subsections.
\subsection{Light transfer}
We first consider the situation that the two variable coupling coefficients, $C_{1}(z)$ and $C_{2}(z)$, are applied in a counterintuitive sequence, such that the coupling coefficient $C_{2}(z)$ precedes the coupling coefficient $C_{1}(z)$ and have the form as follows \cite{Paspalakis2006}
\begin{eqnarray}
\begin{aligned}
C_{1}(z)&=&\kappa \exp[-(z-z_{1})^2/\xi^2],\\
C_{2}(z)&=&\kappa \exp[-(z-z_{2})^2/\xi^2].
\end{aligned}
\end{eqnarray}
\begin{figure}[htbp]
\centering
\includegraphics[width=0.5\textwidth]{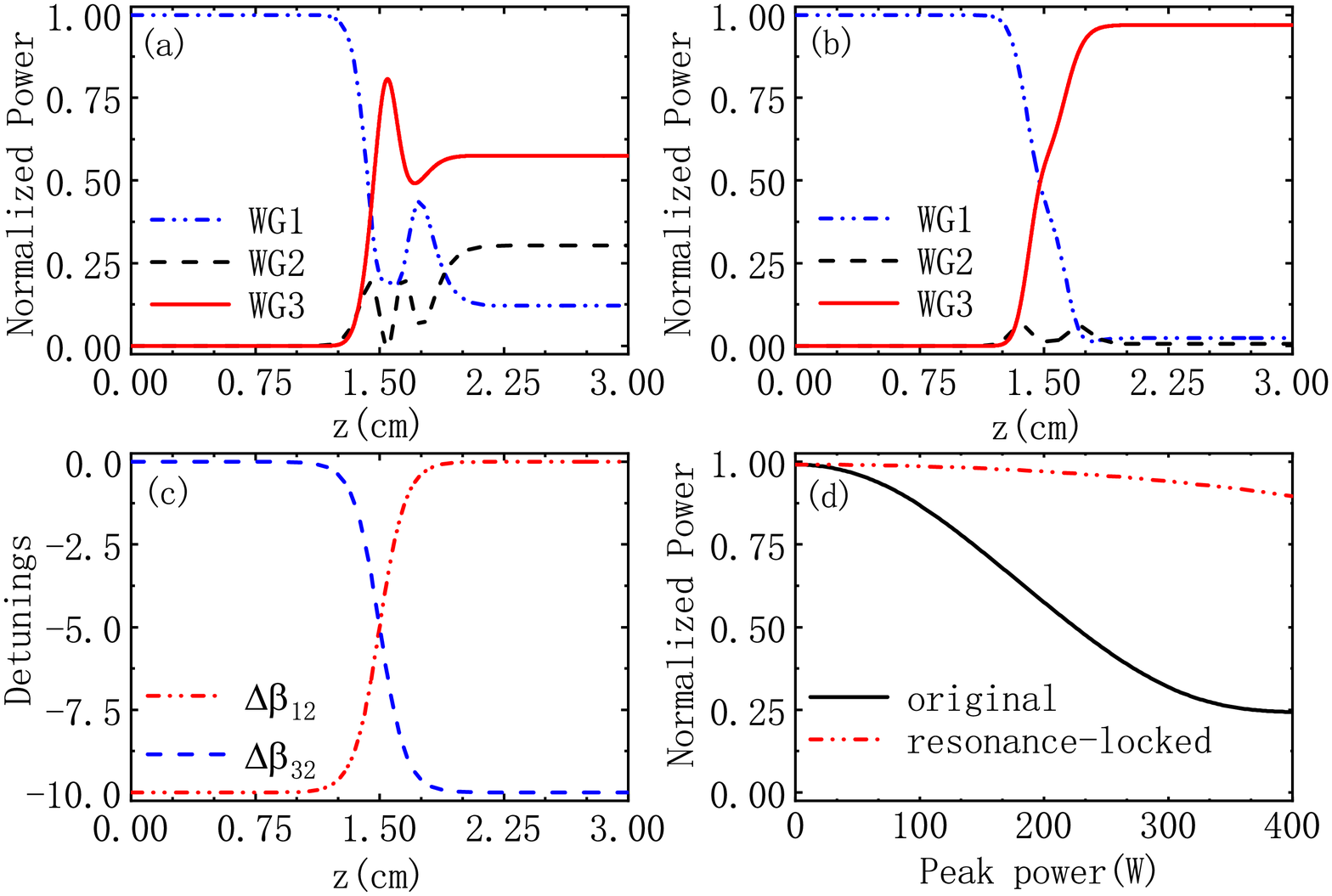}
\caption{Light transfer in the three-waveguide optical system ($P=200W$). The normalized light intensity against the propagation distance $z$ for (a) the original waveguide coupler and (b) the resonance-locked waveguide coupler; (c) The corresponding propagation constants detunings in resonance-locked structure; (d) The output light intensity in WG3 against the input power in original waveguide (solid black line) and resonance-locked coupler (dashed red line). Coupling parameters: $\kappa=15$$cm^{-1}$, $\xi=0.25$$cm^{-1}$, $z_{1}=1.6$cm, $z_{2}=1.4$cm, and $L=3$cm. The propagation constants $\beta_{1}$, $\beta_{2}$ and $\beta_{3}$ are set to a constant i.e., $\beta_{1}=\beta_{2}=\beta_{3}=20$$cm^{-1}$ in original case for convenience.}
\label{figure1}
\end{figure}
Here, $\kappa$ and $\xi$ are the maximum values and the width of the coupling coefficients, respectively. For light transfer, we assume the input as $\mathbf{a}(z_{i})=\lbrack 1,0,0\rbrack^T$ and expect the output final light field distributions among waveguides: $\mathbf{a}(z_{f})=\lbrack 0,0,1\rbrack^T$.
\begin{figure}[htbp]
\centering
\includegraphics[width=0.45\textwidth]{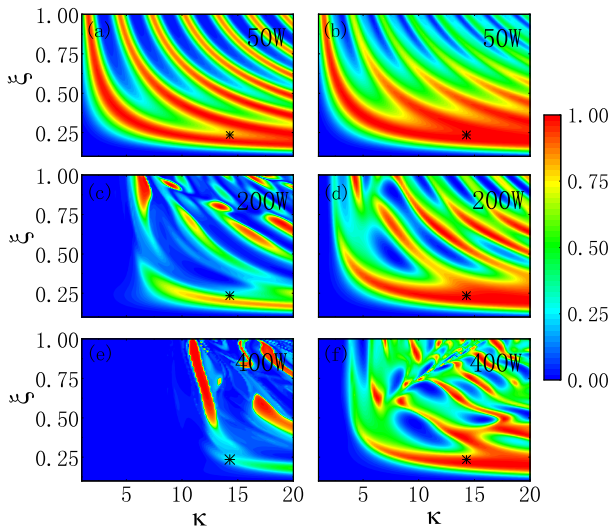}
\caption{The normalized intensity in WG3 at the output port $(z = L)$ as function of the coupling strength parameters $\kappa$ and $\xi$ with different input light power ($P =50W, 200W, 400W$) in original waveguide couplers (left) and resonance-locked couplers (right). The black star spots corresponding to the above coupling parameters in Fig. \ref{figure1}.}
\label{figure2}
\end{figure}
In Fig. \ref{figure1} (a) and (b), we show the normalized intensities in original waveguide coupler and resonance-locked coupler when the input light power $P=200W$. As seen, the original waveguide structure cannot realize the high-fidelity light transfer and normalized light intensities along beam propagation direction are varied irregularly owing to the nonlinear effect. However in the resonance locking case, the power adiabatically leaves WG1 and populates WG3, with very little energy remaining in the intermediate WG2 during the power exchange, in perfect analogy to the linear STIRAP process, which can be understood here the third-order Kerr nonlinearities here are compensated. We also give the shape of the detunings which depend essentially on the population dynamics in Fig. \ref{figure1} (c).  And beyond that, the intensity at the end of the device at WG3 varies with the input optical power is depicted in Fig. \ref{figure1} (d), we note that the fidelity of the light transfer for the resonance-locked coupler is robust against variations of input optical power, while the output intensity in WG3 continue to reduce in original case.

To compare the performance of the proposed resonance-locked coupler to the original one, we depict the contour plots of the light intensity at the end of the device of WG3, as a function of coupling parameters $\kappa$ and $\xi$. We can find that at low input light power ($P=50W$), the output intensity of WG3 is stable in some regions of $\kappa$ and $\xi$, we do not notice any
difference with or without compensation of the Kerr terms. At this time, the influence of nonlinearity on adiabatic evolution can be almost ignored. When the input light power increases ($P=200W,400W$) where Kerr nonlinearities cannot be disregarded, the original waveguide structures are failed to realize adiabatical light transfer, in contrast, resonance locking structure can still achieve high-fidelity light transfer. The small light intensity discrepancies in Fig. \ref{figure2} can be attributed to the fact that the evolution is never perfectly adiabatic, and some non-adiabatic coupling are always present, which limits the efficiency of the STIRAP. In addition, the usual adiabatic condition for linear STIRAP does not hold in the nonlinear case \cite{PhysRevA.103.023307,PhysRevA.78.053410}, thus in some cases, our method cannot eliminate the nonlinearity completely.
\subsection{Light split}
Again we consider light initially is injected in the outer WG1 with $\mathbf{a}(z_{i})=\lbrack 1,0,0\rbrack^T$, the formulation of $C_{1}(z)$ and $C_{2}(z)$ as follows  \cite{PhysRevA.80.013417},
\begin{eqnarray}
\label{15}
\begin{aligned}
C_{1}(z)&=\kappa \sin \varphi,\\
C_{2}(z)&=\kappa \cos \varphi,
\end{aligned}
\end{eqnarray}
where $\varphi(z)=\frac{\pi}{4}/(1+\exp(-(z-L/2)/\omega))$. $\kappa$ is the maximum values of the coupling coefficients and $\omega$ can influence the slope of the coupling constants.
\begin{figure}[htbp]
\centering
\includegraphics[width=0.5\textwidth]{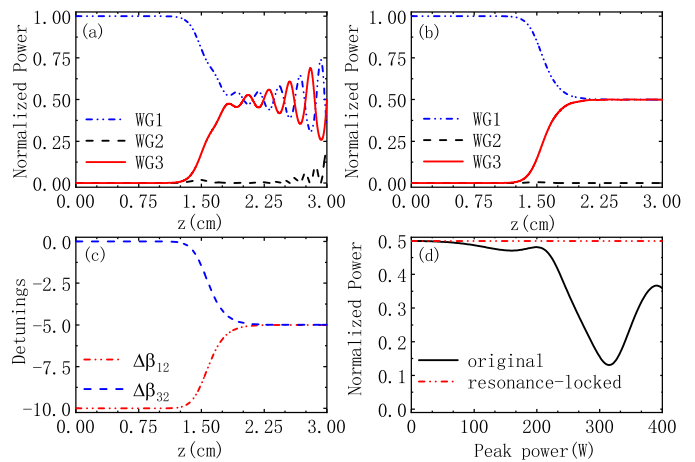}
\caption{Light split in the three-waveguide optical system with the coupling coefficients given by Eq. (\ref{15}). The normalized light intensity against the propagation distance $z$ for (a) the original waveguide coupler and (b) the resonance-locked waveguide coupler; (c) The corresponding propagation constants detunings in resonance-locked structure. In (d) the output light intensity in WG1 for original waveguide (solid black line) and resonance-locked coupler (dashed red line) is shown. The parameters for this figure are $\kappa=25$$cm^{-1}$, $\omega=0.25$$cm^{-1}$, $P=200W$, and $L=3$cm.}
\label{figure3}
\end{figure}
\begin{figure}[htbp]
\centering
\includegraphics[width=0.45\textwidth]{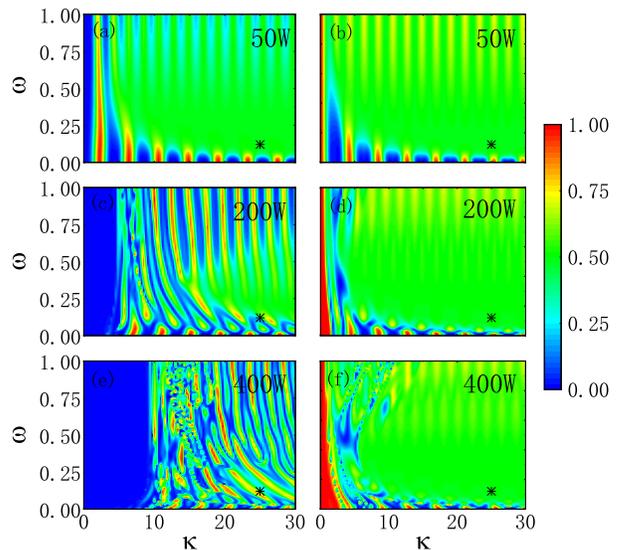}
\caption{As the coupling strength parameters $\kappa$ and $\omega$ vary, the normalized intensity in WG1 at the output port $(z = L)$ with different input optical power in original waveguide couplers (left) and resonance-locked couplers (right) is shown.}
\label{figure4}
\end{figure}
Rather than light transfer, the goal of the light split is to divide equally light intensity in WG1 and WG3, the WG2 almost not excited during the optical evolution. In original waveguide coupler, as the input light power $P=200W$, one cannot obtain a complete split in the presence of three-order nonlinearities which we can see from Fig. \ref{figure3} (a). The nonlinear waveguide system with three-order nonlinearities demonstrate noncontrollable exactly in the sense that such nonlinearities prevent reaching the target state exactly \cite{PhysRevA.88.063622,PhysRevA.102.052203}. However, in resonance-locked scheme, light split can be realized perfectly. The issue of robustness of the output intensity of WG1 with respect to the input light power is further numerically analyzed in Fig. \ref{figure3} (d). As mentioned above, the output light intensity in WG1 for the resonance-locked case is independent of input light power. We also test the robustness of our resonance-locked couplers with respect to coupling parameters $\kappa$ and $\omega$ in Fig. \ref{figure4}, which leads to the same conclusion in light transfer case. As the nonlinear interaction grows ($P=200W, 400W$), non-adiabatic oscillations are significantly strengthened and the light intensity is reduced dramatically, while it is relatively stable in resonance-locked case.
\subsection{Light return}
\begin{figure}[htbp]
\centering
\includegraphics[width=0.5\textwidth]{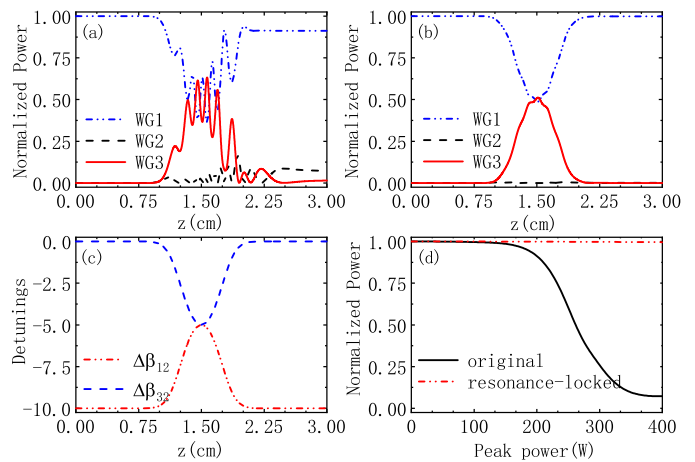}
\caption{Light return in the three-waveguide optical system. The normalized light intensity against the propagation distance $z$ for (a) the original waveguide coupler and (b) the resonance-locked waveguide coupler; (c) The corresponding propagation constants detunings in resonance-locked structure. In (d) the output light intensity in WG1 for original waveguide (solid black line) and resonance-locked coupler (dashed red line) is shown. In this example, we establish the parameters as $\kappa=40$$cm^{-1}$, $\xi_{1}=0.3$$cm^{-1}$, $\xi_{2}=0.6$$cm^{-1}$, $P=200W$, and $L=3$cm.}
\label{figure5}
\end{figure}

Once more we consider the input as $\mathbf{a}(z_{i})=\lbrack 1,0,0\rbrack^T$ and choose the coupling coefficients to be Gaussian type of the form \cite{Paspalakis2006}
\begin{eqnarray}
\begin{aligned}
C_{1}(z)&=&\kappa \exp[-(z-L/2)^2/\xi_{1}^2],\\
C_{2}(z)&=&\kappa \exp[-(z-L/2)^2/\xi_{2}^2],
\end{aligned}
\end{eqnarray}
where the $\kappa$ denotes the maximum values of the coupling coefficients and the parameters $\xi_{1}$ and $\xi_{2}$ represent the the widths of the coupling coefficients. It is well known that the system will regularly undergo the complete population return (CPR) with $\mathbf{a}(z_{f})=\lbrack 1,0,0\rbrack^T$ when excited by a resonant coupling \cite{Vitanov1995}, i.e., the process is reciprocal. The coupling coefficients must only be sufficiently wide and overlap significantly in distance so that adiabatic evolution is ensured \cite{PhysRevA.103.053705}. Finally results of complete light return to the initially excited waveguide WG1 is shown in Fig. \ref{figure5}. During the propagation, the other two waveguides WG1 and WG3 take part in the coupling in Fig. \ref{figure5} (a). The special feature is the appearance of pronounced oscillations due to the non-adiabatic transition between the different eigenstates \cite{PhysRevA.73.013617}. We, however, notice that our resonance-locked waveguide coupler could obtain the almost complete population return even in the presence of Kerr terms in Fig. \ref{figure5} (b).
\begin{figure}[htbp]
\centering
\includegraphics[width=0.45\textwidth]{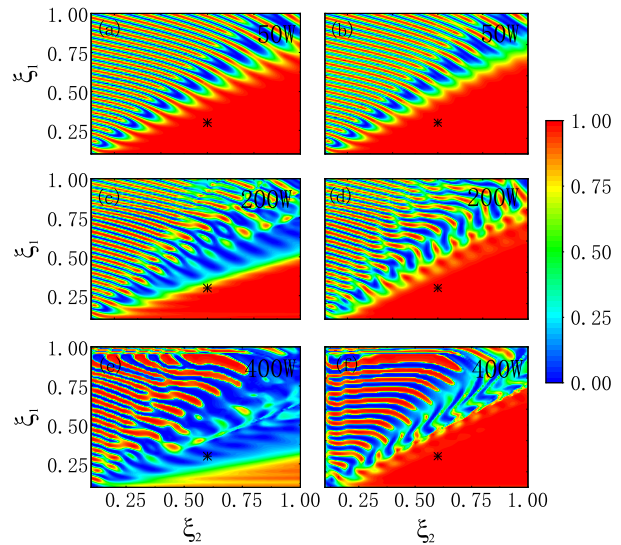}
\caption{The output light intensity in WG1 for original waveguide (left) and resonance-locked coupler (right) as a function of the widths of the coupling coefficients $\xi_{1}$ and $\xi_{2}$ with the different input optical power is depicted.}
\label{figure6}
\end{figure}
We continue to show the superiority of the resonance-locked waveguide couplers as compared to the original waveguide couplers by examining the light intensity return to WG1, as a function of the widths of the coupling coefficients $\xi_{1}$ and $\xi_{2}$ and the maximum value of the coupling coefficients is set to a constant $\kappa=40$$cm^{-1}$ for simplicity. Apart from that, it is shown that the output light intensity in WG1 is quite sensitive to the input light power in original case, indicating the nonadiabaticity of the light evolution in Fig. \ref{figure5} (d). We note that our resonance-locked waveguide coupler has a higher tolerance to the fluctuations in various parameters such as the width of Gaussian pulses and the input light power than original case in Fig. \ref{figure6}. The most intriguing property presented here comes from the three-order nonlinearity, as the nonlinearity gets stronger (corresponding to $P =200W,400W$), the adiabatic evolution is greatly disrupted and we observe a rapid fall of the robustness if one does not compensate the Kerr terms with the detunings.

We note that the coupling coefficients here need not be of a specific shape nor their parameters need to obey specific conditions, in contrast to the conventional nonlinear coupling approach. The choice of coefficients in the above three examples were made only for convenience.
\section{Conclusions}\label{sec:4}
In conclusion, we have theoretically employed the resonance-locked inverse method based on STIRAP for light evolution in nonlinear waveguide couplers. The presence of third-order Kerr nonlinearities which critically depends on the light intensity and the excitation power, impairs the efficiency of the STIRAP. It was shown that, upon an appropriate choice of the longitudinal evolution of the propagation constants detunings, the effect introduced by the third-order nonlinearities have been dynamically compensated, thus, the above nonlinear system exhibits like the linear regime subjected to the adiabatic passage process. Both the theoretical analysis and numerical calculations show that high-fidelity and robust light transfer, light split, and light return can be observed in the our resonance-locked waveguide structures in contrast to the case of the conventional nonlinear couplers. In addition, the proposed scheme is not constrained by the need for having exact system parameters and is not limited by Kerr nonlinearities parameters which also promises myriad possibilities in designing robust nonlinear waveguide structures such as beam splitters, switches, and directional couplers. Therefore, the light propagation in nonlinear waveguide couplers via the resonance-locked inverse based on STIRAP protocol is intriguing and vigorous topic in circuits and communications, magneto-optic data storage and many other practical fields.

The key point of implementation of resonance-locked inverse method based on STIRAP is the space-dependent propagation constants detunings in the waveguide coupler fabrication, which can be approximately realized by modifying the refractive index of waveguides along the propagation direction. Experimentally, the combination of the local illumination of the control beam and the electric field applied to the Sr$_{x}$Ba$_{1-x}$Nb$_{2}$O$_{6}$ (SBN) crystal \cite{PhysRevA.103.053705,PhysRevA.95.023811,Gorram2009} with $x = 0.61$ could lead to the above expectation, which also reflect the technical feasibility \cite{math8071128,Zhang2022}. It should
be noted here that, different from the linear waveguide systems works \cite{PhysRevA.103.053705,PhysRevA.95.023811,Gorram2009}, we consider the appropriate choices of the longitudinally varying detunings could fully compensate the nonlinear effect to realize the high-fidelity light propagation even in the nonlinear regime. Another distinguishing feature is that others discuss the spatial longitudinally varying detuning between the propagation constants of the waveguides could provide with more freedom than $z$-independent detuning to manipulate light evolution \cite{PhysRevA.103.053705,PhysRevA.95.023811}. Moreover, we believe this study may open new possibilities of exploiting the resonance-locked inverse scheme based on STIRAP for various applications in integrated optics.
\section*{Acknowledgments}
The work is supported by the National Natural Science Foundation of China (Grant No. 12075193).
\bibliography{reference}
\end{document}